# Lower leg compression and its biomechanical effects on the soft tissues of the leg

Fanette Chassagne[1,2], Pierre Badel[1], Jérôme Molimard[1]

[1] *Mines Saint-Etienne, Univ Lyon, Univ Jean Monnet, INSERM, U 1059 Sainbiose, Centre CIS, Saint-Etienne France*

[2] *Thuasne, Saint-Etienne, France*

**Abstract (100-150 mots)**

Elastic compression of the lower leg is the traditional preventive and curative treatment of venous insufficiency. After presenting the medical strategies related to compression therapies, this chapter develops current advances in clinics as well as in engineering and outline the more important knowledge arising from this review. Compression hosiery acts by providing pressure to the leg. Pressure generation using socks mainly depends on the stiffness and the size of the socks, the size of the leg, but also the leg morphology. In the case of bandages, the role of friction must be outlined, as the main factor in maintaining the bandage wrapped around the leg, but also as an influencing factor in pressure generation. Besides generated pressure, response of superficial veins to compression also depends on the vein size and the fat stiffness. But mechanical assessments should not mask the importance of other factors such as muscular contraction or nurse formation. An important impact of these results would be head towards an improved personalization of compression treatment.

## 1 - Introduction

The role of the venous system is to flow the blood back to the heart. In the lower leg, in standing position, blood flow must counteract gravity. For this, several mechanisms contribute to venous return: the sympathetic nervous activity, the calf pump and the valves in the lower leg veins. If one or several of these mechanisms become deficient, blood stagnation and reflux may occur and lead to venous diseases.
Elastic compression of the lower leg is the traditional preventive and curative treatment of venous insufficiency. This compression, whose clinical efficacy is widely admitted, can either be performed thanks to stockings or bandages. Compression stockings are made of elastic fabric and are sized to the patient leg. They are prescribed for the treatment of non-acute stages of venous insufficiency and can be put on by the patient or a non-trained person. Bandages consist in a stretched fabric wrapped around the leg and are mainly used to treat the most severe pathologies, such as venous ulcers. As they are not sized to the patient's leg, they are also recommended for the first stages of the treatment, when the leg swelling is reduced rapidly. Moreover, they are prescribed when the patient's pathology prevents the use of stockings (for knee arthroplasty for instance). However, bandages must be applied by experienced care givers and this treatment modality is very operator dependent.

Both medical devices apply a pressure onto the skin called interface pressure. The basic principle of compression therapy may seem simple: the pressure applied onto the skin is transmitted through soft tissues (muscle and adipose tissues - Figure 1) to the veins. This pressure tends to reduce the diameter of the vein and consequently the blood volume in the leg. However, though the efficacy of this treatment is admitted, the therapeutic objectives are not always reached, underlying the fact that the mechanisms of compression therapy are not yet fully understood.

A first step towards improved understanding is to investigate how interface pressure is generated, then how it is transmitted to the veins and how this pressure will enhance venous return. *In vivo* measurements of interface pressure are a common clinical approach but investigating pressure transmission, for instance, is more challenging. For this, engineering approaches such as numerical simulation may represent promising tools. Numerical simulation can provide *in situ* information that is very difficult or impossible to measure and could result in the prediction of patient-specific outcomes of the treatment. After a description of the medical management of compression treatment, this chapter introduces the current advances in both clinical and engineering approaches designed to study the biomechanical effects of compression therapy. Perspectives and current questions are addressed at the end of the chapter.

## 2 - Medical strategy

### 2.1 - Venous return & chronic venous insufficiency

Even though the pathogenesis of venous insufficiency is not fully understood (Santler and Goerge 2017), this pathology occurs when venous return becomes less competent. The role of the venous system is to flow deoxygenated blood from the lower extremities to the heart. In standing position, because of gravity, the intravenous pressure increases from the upper leg to the lower leg (B. Partsch and Partsch 2005) and can reach 75 mmHg (Gardon-Mollard and Ramelet 2008). Thus, venous return must play against pressure gradient and gravity (Meissner et al. 2007).

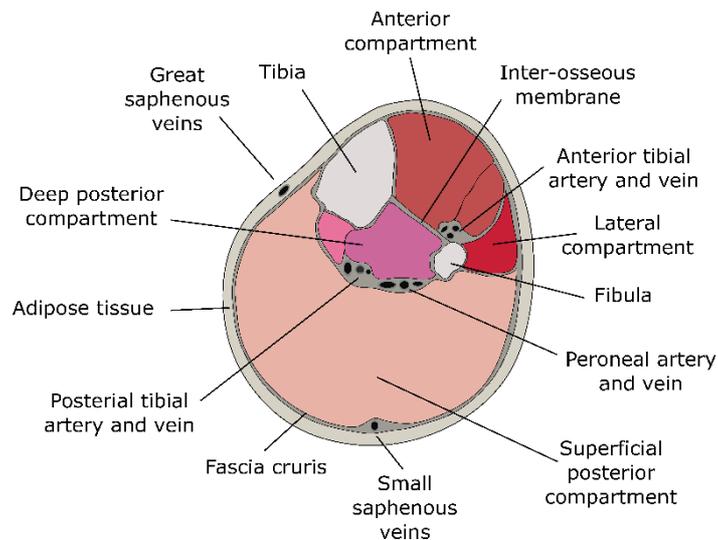
Figure 1: Cross section of the lower leg - adapted from (Braus and Elze 1921)

For this, three main mechanisms compose venous return:
- *the sympathetic activity on venous return*; The venous wall is composed of smooth muscular cells which can contract under the effect of the sympathetic nervous system. When the flow is too low, the cells contract to enhance venous return and counteract gravity (Marieb and Hoehn 2010).
- *the contraction of the calf muscles and the vein valves*; While contracting, the calf muscles compress deep veins. These veins have valves which prevent backflow. The compression of veins by their surrounding muscles results in an increase in intravenous pressure and the opening of the valves. Blood will be propelled to the heart and valves will close again when the pressure decreases (Figure 2). The calf pump provides about 50% of the energy required for venous return (Marieb and Hoehn 2010).

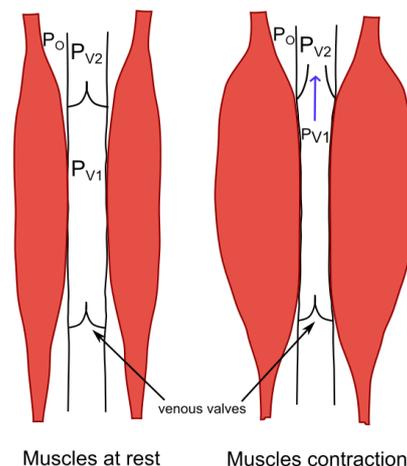
Figure 2: Illustration of the calf pump and deep veins valves

A lower efficiency of venous return will lead to blood stasis, reflux and an increased intravenous pressure. This pathology is called chronic venous insufficiency and ranges from 'heavy leg' feelings to venous ulcers. The most common conservative and preventive treatment for this pathology is compression therapy, which can be performed thanks to stockings, bandages or intermittent pneumatic compression devices.

The main objective of compression therapy is to reduce the difference in pressure between the inside and the outside of the venous wall (Gardon-Mollard and Ramelet 2008) – also called transmural pressure - in order to reduce the caliber of veins and facilitate venous return. The action mechanism of compression consists in applying a pressure onto the leg, which is called interface pressure. This external pressure is then transmitted through soft tissues (adipose and muscle tissues - Figure 1) to the veins (P.-Y. Rohan et al. 2015; B. Partsch and Partsch 2005).

Though it is widely admitted that compression enhances venous return, the action mechanisms of compression therapy are not fully understood (Lattimer et al. 2014) and neither is their relative influence. Nevertheless, irrespectively of compression modality, interface pressure remains the key aspect of the treatment.

## 2.2 - Interface pressure as a therapeutic dosage

Interface pressure applied by compression devices should be adapted to each patient and his/her pathology (Vicaretti 2010). However, the level of pressure recommended for each pathology is not clearly defined. It was observed that even low pressure (about 20 mmHg) can improve venous pumping (Mosti and Partsch 2010a). Increasing interface pressure (up to 60 mmHg) would result in a larger improvement of venous pumping (Mosti and Partsch 2010a) and was proven to have a better therapeutic effect on ulcers (Milic et al. 2010). Higher interface pressure can result in veins narrowing. This would lead to a reduction of venous reflux (H. Partsch, Menzinger, and Mostbeck 1999) and venous hypertension (H. Partsch 1984) and to an increase of calf pump function (Mosti, Mattaliano, and Partsch 2008). However, if the applied pressure is too high, the treatment becomes less bearable for patients and can lead to complications such as pressure ulcers (Mosti and Partsch 2010a). For the best compromise between treatment observance and efficacy, it is often recommended to apply the highest pressure the patient can tolerate.

In order to standardize the quantification of compression therapeutic dosage, standard measurement points have been defined for *in vivo* interface pressure measurement (H. Partsch et al. 2006) (Figure 3 - A):
- Measurement point B: right above the ankle where the leg circumference is the smallest
- Measurement point B1: when the Achille's tendon turns into the gastrocnemius muscles
- Measurement point C: at the largest circumference of the calf.

Measurement point B is the reference point for the design of textile compression devices and is also the location where most ulcers occur (Callam et al. 1987). Lately, measurement point B1 was considered as the point of interest for *in vivo* pressure measurements (H. Partsch et al. 2006). In this area, the change in leg circumference from supine to standing position (and while walking) is very large. Thus, interface pressure variations in this area are more important (F. Chassagne et al. 2017).

Yet, interface pressure is also influenced by the patient's morphology. Compression stockings are sized to the calf circumference, within the aim at applying similar pressures for all morphologies. To design compression stockings, manufacturers commonly use wooden legs of different sizes (the so called Hohenstein legs) to assess the pressure applied by a stocking at measurement point B (Afnor 1986). The target pressures are divided into "compression classes" which vary among countries (Table 1).

| Compression class | USA | UK | France | Germany |
|---|---|---|---|---|
| I | 15-20 | ≤ 20 | 10-15 | 18-21 |
| II | 20-30 | 21-30 | 15-20 | 23-32 |
| III | 30-40 | 31-40 | 20-36 | 34-46 |
| IV | > 40 | 41-60 | > 36 | > 49 |

Table 1: Pressures (in mmHg) at measurement point B for different compression classes and countries (Rabe et al. 2008)

On the other hand, compression bandages are applied with the same stretch for any patients' morphology. The same bandage being applied on different patients' morphology will result in different interface pressures (F. Chassagne et al. 2015).
Also, pressure may vary between a passive muscular state (relaxed calf muscles) and an active muscular state (active calf muscles, while walking for instance) (Dissemond et al. 2016). These pressures are called resting and working pressures. Thus, interface pressure is also influenced by body position (H. Partsch 2005b).
To reach the appropriate interface pressure, there are several ways of modulating the compression treatment. An additional compression stocking can be worn on top of a stocking to increase interface pressure (Cornu-Thenard et al. 2007). For compression bandages, varying the application technique (either the wrapping pattern or the bandage overlapping) would have an impact on the pressure (Coull, Tolson, and McIntosh 2006). Also, it is possible to combine 2 to 4 bandages with different mechanical properties to modify the applied interface pressure and its variations with body position (Hanna, Bohbot, and Connolly 2008).

### 2.3 - Progressivity vs degressivity
Most often, compression stockings or bandages apply a degressive profile of pressure: the applied pressure decreases from the ankle to the knee (Figure 4 – A – left). This profile is mainly the consequence of the conical shape of the leg: the radius of curvature is smaller at the ankle than at the larger circumference of the calf (Figure 3 - B). Also, such pressure profile tends to counteract the increase of intravenous pressure from the knee to the ankle due to gravity. The therapeutic efficacy of degressive compression profile is widely admitted (Amsler, Willenberg, and Blättler 2009).

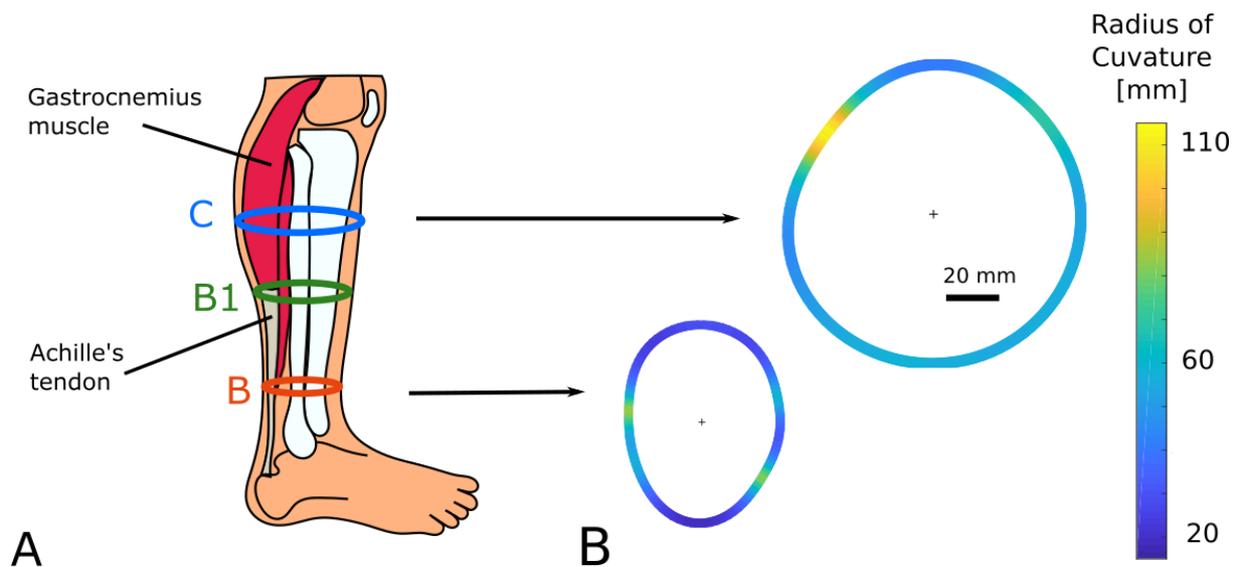

Figure 3: A- Location of standard measurement points; B- Illustration of the variation of local radii along the leg

Lately, another type of compression profile was tested, called progressive. In the case of a progressive compression device, the pressure applied at the largest circumference of the calf (measurement point C) is higher than that at the ankle (Figure 4 – A – right). The therapeutic effect of progressive compression was assessed either with stockings (Mosti and Partsch 2011) or bandages (Mosti and Partsch 2012). For the least severe forms of the pathologies, the efficacy of progressive compression was found to be equivalent to degressive compression (Couzan et al. 2009). Whether progressive compression is performed with stockings (Mosti and Partsch 2011) or bandages (Mosti, Iabichella, and Partsch 2012), the positive impact on venous return (venous ejection fraction) was found to be higher than with degressive compression. However, these outcomes result from short term experiments and the long term effects of progressive compression still needs to be investigated (Shepherd 2016).

## 2.4 - Elasticity and stiffness

An important property of compression therapy with bandages is elasticity. In the case of bandages, elasticity is not defined as the capacity of the material to return to its original shape after being stretched (Clark 2003). Instead, bandages are divided in three groups depending of the bandage stretch under a given applied force (10 N/cm - (DIN 61632 2009)) (Table 2).

|  | Inelastic |  | Elastic |
|---|---|---|---|
|  | Rigid | Short-stretch | Long-stretch |
| Maximal stretch at 10 N/cm bandage width | 0-10 | 10-100 | >100 |

Table 2: Classification of compression bandages

Usually, elasticity and extensibility are connected (H. Partsch et al. 2008): long-stretch bandages are associated with so-called "elastic" properties and short-stretch bandages with so-called "inelastic" properties. The main clinical difference between bandages with different mechanical properties is the pressure variation from supine to standing position (H. Partsch 2005b). Compression bandages are applied on the lower leg while the patient is lying in supine position. From supine to standing position, the leg volume and the calf mechanical properties change. Elastic bandages will tend to adapt to these changes, which will result in a low pressure increase from supine to standing position (Figure 4 - B). For inelastic bandages, the pressure increase is higher (Clark 2003) (H. Partsch 2005b).

Besides mechanical tensile testing, a clinical method allows assessing the mechanical properties of a bandage: the *Static* Stiffness Index (SSI). The SSI is the pressure increase from supine to standing position (H. Partsch 2005a). A bandage is considered as inelastic if the SSI is higher than 10 mmHg (H. Partsch 2005b).

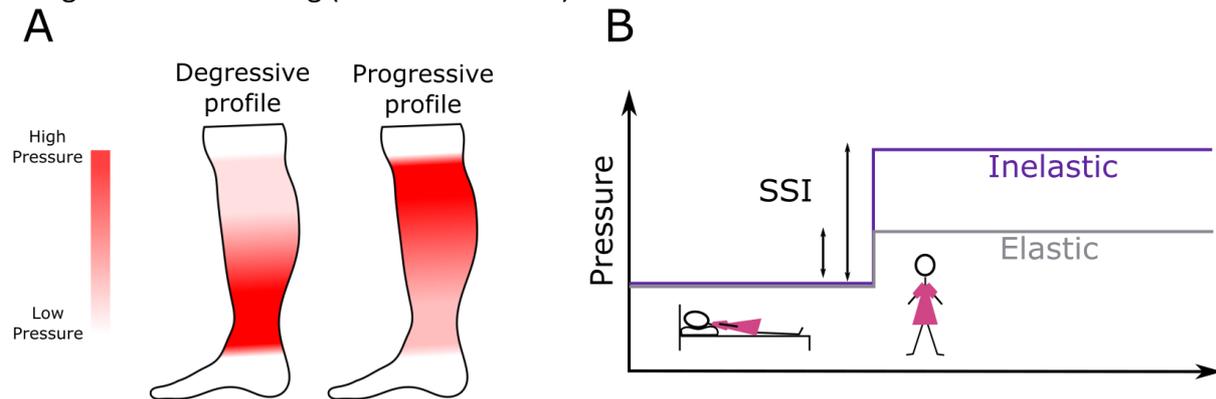

**Figure 4: A- Illustration of a degressive and a progressive pressure profile; B- SSI for an elastic and an inelastic bandage**

The difference in elasticity is the cornerstone of the difference between compression and contention. *Compression* consists in the effects induced on a leg by an elastic bandage (active), whereas *Contention* is the passive action of an inelastic fabric on the leg (Gardon-Mollard and Ramelet 2008). The inelastic fabric will oppose to a change in leg morphology (from supine to standing position for instance), whereas elastic fabric will comply with such change and apply quasi-constant pressure on the leg.

Different bandage properties will result in different clinical outcomes. It was shown that the elasticity of the bandage may have a positive impact on ulcers healing (Danielsen, Madsen, and Henriksen 1998). Other studies have observed a better improvement of venous return with short-stretch bandages compared to long-stretch bandages (Mosti, Mattaliano, and Partsch 2008) (H. Partsch, Menzinger, and Mostbeck 1999) .

## 2.5 - Different schools of compression

The diversity of modalities for compression therapy is reflected by the different schools of compression within countries. For instance in France, elastic compression is commonly prescribed: mainly elastic stockings and elastic bandages (Gardon-Mollard and Ramelet 2008). In Germany and Austria, the standard treatment is also performed with stockings and bandages. However, inelastic compression is usually prescribed. In England, the four-layer bandage is the standard treatment for venous ulcers (Blair et al. 1988). This bandage is composed of a wool bandage (padding layer), a creped bandage, an elastic bandage and a cohesive bandage (Moffatt 2002). In the Netherlands, the treatment of venous ulcers is mainly performed with inelastic compression or intermittent pneumatic compression devices.

Even though the kind of compression therapy may differ among countries, it is most often the standard treatment for venous insufficiency. Nevertheless, in the USA, compression stockings are sparingly prescribed for chronic venous insufficiency or after sclerotherapy, and compression bandages are never prescribed.

The prescription of the appropriate compression is a real challenge. It must be adapted to the patient and its pathology while being the best compromise between a high interface pressure and comfort for the patient (and compliance to the treatment).
The large discrepancies among treatment modalities and traditions underline the need of a better understanding of the mechanisms of compression therapy, which was addressed through clinical and engineering studies.

3 - **Clinical approach**

**3.1 - Pressure measurement**
*Pressure measurement systems*
The main effect of elastic compression is the pressure applied on the skin that induces a change in depth, at the levels of tissues and veins. Therefore, the pressure applied onto the leg has been measured and/or estimated in many studies as an efficacy indicator.

Pressure sensors are based on four main technologies: pneumatic, electro-pneumatic, capacitive or resistive, corresponding to well-known commercial products, for example respectively, PicoPress (Neots Cambridgeshire, United Kingdom), Salzmann (Boston MA, USA), X-Sensor (St. Gallen, Switzerland) and finally Tekscan (Calgary Alberta, Canada) or Novel (Munich, Germany). Resistive or capacitive sensors may be assembled into a structure that enables the pressure to be measured at several points simultaneously; this structure is often called a pressure mapping system.
In the medical field, many authors have used pressure measurement to evaluate medical devices. They have been employed to prevent bedsores (Fenety, Putnam, and Walker 2000; Hamanami, Tokuhiro, and Inoue 2004; Reenalda et al. 2009; Rithalia, Heath, and Gonsalkorale 2000; Shelton and Lott 2003; Tam et al. 2003; Vaisbuch, Meyer, and Weiss 2000)), wheelchair ulcers (Meffre, Gehin, and Dittmar 2007), in the study of rigid or soft braces (Aubin et al. 2007; van den Hout et al. 2002; Rébecca Bonnaire et al. 2014) or more recently for ulcer prevention in diabetic patients (Perrier et al. 2014; Bus et al. 2016) or even in the study of bandage on animals (Taroni et al. 2015). Yet, most work has been developed to measure the interface pressure of compression stockings or compression bandages (Candy, Cecilia, and Ping 2010; Damstra, Brouwer, and Partsch 2008; Dumbleton et al. 2009; Hafner et al. 2000; Kumar, Das, and Alagirusamy 2013; H. Partsch, Partsch, and Braun 2006; Troynikov et al. 2013; Wheeler et al. 2011).
For a given study, the choice of the right pressure sensor has to made with care. Pneumatic and electro-pneumatic sensors are easy to use and rather low cost but the quality of measurements depends on the user. These sensors give a single measurement on a large area compared to the expected pressure spatial variations on the leg. Also, these sensors have a low (i.e. good) resolution that is well adapted to the pressure range to be measured on a leg. Capacitive sensors usually have a low measurement error but are more expensive and their drift makes them more relevant for dynamic measurements than static (most usual) measurements. The latest resistive sensors have a good resolution and a reasonable price compared to capacitive ones. However, their sensitivity to surface curvature has been pointed out, even if piezo-resistive ones seem to be less affected.

From this brief overview of existing technologies, it is shown that the choice of a pressure measurement device should be driven by the expected resolution, measuring range, acceptable error and cost, but also by the size and the curvature of the measured area. Note that sensors with a curvature fitting the curvature of the zone to be measured have even been specifically designed, for example for rigid bracing (Lou et al. 2008) or for prosthetic sockets (Wheeler et al. 2011). Last, it should be mentioned that in compression therapies, most of clinical or experimental work has been done using pointwise pneumatic or electro-pneumatic sensors (H. Partsch and Mosti 2010). An overview of these studies is presented below.

**3.2 - Clinical studies**
Clinical studies related to elastic compression address different health problems. They have been applied to venous ulcers, varicose veins, deep vein thrombosis, post-surgery, post-sclerotherapy or post-phlebitic legs. The use of compression hosiery opened the way to various questions: the compression level, the impact of the hosiery design on the desired level of compression in particular with time, the representativeness of phantom legs, the patient compliance…

Different classifications are used in compression hosiery (Table 1). They are based on local practices (Great Britain, France, Germany, USA) and define compression levels and pressure degressivity. As a logical consequence, several studies investigated the pressure effectively applied on the leg. Reich-Schupke (Reich-Schupke et al. 2009) showed that 32% were out of the target, and among them a quarter above, which could result in clinical complications. Partsch established compression pressure to achieve vein closure on small saphenous vein and concluded that this pressure varies from 15 mmHg in supine position to 70 mmHg in standing position (B. Partsch and Partsch 2005).

The quality of the hosiery has been questioned, either at the design level or at the clinical one. For example, Kumar (Kumar, Das, and Alagirusamy 2013) or Troynikov (Troynikov et al. 2013) studied the influence of different materials and knitting patterns on the pressure evolution with time; Ma (Ma et al. 2015) compared the efficacy of different brands, among them low-cost ones, without significant difference between low-cost and others. It is worth noting anyway that these studies used universal testing machine and no clinical verification was provided. In a dedicated study, Partsch (H. Partsch, Partsch, and Braun 2006) concluded on the good correlation between *in-vitro* and *in-vivo* measurements.

Variations of pressure were also investigated over time (Candy, Cecilia, and Ping 2010; Damstra, Brouwer, and Partsch 2008), or with patient activity (H. Partsch 2016). Viscoelasticity effects have been widely demonstrated and can explain part of the pressure loss observed over time. However, Damstra (Damstra, Brouwer, and Partsch 2008) suggested that the pressure loss is mainly due to the leg volume variation coupled with the strong non-linear elastic behavior of the so-called 'inelastic' textiles. In the same way, Partsch (H. Partsch 2016) established the interaction between the stocking and muscular pump function.

In a review focused on uncomplicated varicose veins, Palfreyman outlined the variety of indicators for the blood return (Palfreyman and Michaels 2009). Authors have measured foot or leg volume variations, venous reflux, skin blood flow, blood volume or pressure variation… Despite standards already reported, the adequate compression level is still questioning. Even if higher pressures usually lead to better results, the patient compliance is low in this case. Only 37% of patients in initial clinical evaluation maintained their treatment (50% for the one who had history with deep vein thrombosis). As a matter of fact, even if a link between pressure and comfort is intuitive, no correlation was sought.

The case of bandages is specific because these devices fit better to any leg shape and are suitable for severe cases. However, they are more dependent on the practitioner's experience and show a wider variability in the applied pressure. Studies on this type of compression device are less frequent than for stockings or socks. Most of the measurement campaigns which were performed on men and women did not take the sex difference into account (Mosti and Partsch 2013, 2010b), however the leg morphology has an influence on the applied pressure and varies from a subject to another, maybe even more especially from a female subject to a male subject. Other groups investigated the impact of the application technique on the interface pressure (Coull, Tolson, and McIntosh 2006) and they also measured the stretch of the applied bandage. However, this study was carried out for only one bandage type. Other studies were focused on the influence of bandage mechanical properties and position (supine, standing, sitting) on the interface pressure (Danielsen et al. 1998; Hirai et al. 2009; Benigni et al. 2008; Rimaud, Convert, and Calmels 2014). Of particular interest, Rimaud (Rimaud, Convert, and Calmels 2014) concluded on the high variability of interface pressure between subjects for some compression bandages (inelastic fabrics, figure-of-eight method). Yet, this variability can be reduced by training nurses, as noted by Hafner who showed that no bandage exceeded the vein collapse pressure after training of the care giver (Hafner et al. 2000). Recently, Chassagne added the measurement of bandage stretch in a clinical study (F. Chassagne et al. 2015). The objective of this study was to perform a complete campaign including bandage stretch and pressure measurements in order to evaluate the influence of subject morphology, position, bandage mechanical properties and application technique on interface pressure applied by elastic compression bandage. It revealed a very strong correlation between the applied pressure and the bandage mechanical properties but also between the pressure and the application technique, hence the need to control the bandage stretch and the application technique to control interface pressure.

Despite the quantity of literature available on compression hosiery and its capacity in delivering the appropriate therapeutic dosage of pressure, there is still a lack of understanding that have been expressed very recently by Hageman (Hageman et al. 2018): "In short, a fundamental understanding of the connection between compression therapy, modes of achieving compression, disease onset, and disease progression is needed". Indeed, pressure measurements are usually pointwise measurements on the surface of the leg. Other methodologies measuring blood return are useful global information, but none could describe a mechanism of action from the skin to the veins.

### 3.3- Modeling pressure using the law of Laplace

For years, using the law of Laplace has been a classical way to estimate the pressure applied by a compression device on the body (Thomas 2003). It is used in particular to qualify hosiery in the French standard (Afnor 1986) by converting the textile stiffness into pressure. The law of Laplace was proposed by both Pierre Simon de Laplace and Thomas Young in 1805 and describes the pressure difference between two fluids (for example: droplets in emulsions, capillaries). It relates this pressure difference to the surface tension and the radius of curvature. It has been applied later to fluid/membrane interface and is now well-established in medicine: alveoli of the lungs as well as blood vessels can be described by the law of Laplace. It was applied to compression therapies in the 1980's (Thomas 2003). According to the law of Laplace, the pressure $p$ applied on the leg by compression hosiery is linked to the radius of curvature $R$ of the leg and the tensile force per unit width $T$ of fabrics as following:

$$p = \frac{T_1}{R_1} + \frac{T_2}{R_2}, T_1, T_2 \geq 0, R_1, R_2 > 0 \quad (1)$$

with direction 1 and 2 being the principal radii of curvature.

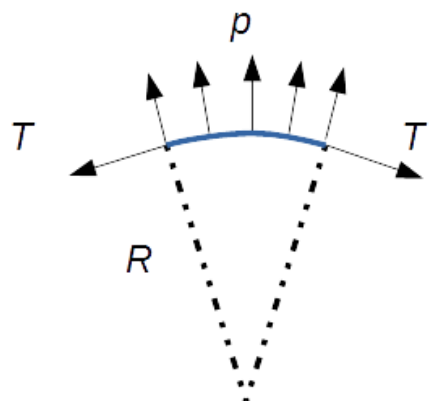

Figure 5: Law of Laplace

When getting into details, many limitations affect the law of Laplace. First, the pressure estimated by equation (1) is valid only for convex surfaces. If the curvature is concave, the pressure is zero as the textile would not be in contact with the body. Second, it is assumed that the two directions of traction correspond to the two main curvature radii; this assumption is reasonable for devices applied on a leg, which is mostly cylindrical, with one prevailing direction. Third, the textile is considered as a membrane, with no compression or bending behavior. Last, the law of Laplace assumes that the textile is sliding on the body without friction. In the specific case of leg compression, this assumption is not fulfilled, otherwise the garment would fall on the patient's ankle as Chassagne (Fanette Chassagne et al. 2018) showed.

Because of these limitations, authors tried to modify the law of Laplace to take extra phenomena into account. The most classical presentation includes the number of layers in a bandage (Schuren and Mohr 2008). Many authors discussed the law of Laplace's validity by comparing with experiments on phantom legs or on patients with contradictory conclusions (Aghajani, Jeddi, and Tehran 2011). Even if no author directly rejected the law of Laplace, there is still a lack in this analytical description to get predictions with enough precision. According to Chassagne (F. Chassagne et al. 2016), leg deformation – especially during wrapping of bandages – and adhesion are the two main factors for this. A first attempt to overcome this limitation using semi-analytical approach has been proposed recently (R.

Bonnaire et al. 2016) but the only complete way to address the problem is numerical modelling, which is detailed in the following section.

**4 - Engineering leg compression**
As presented in the previous section, a large amount of clinical studies focused on characterizing the interface pressure generated by compression textiles. Still, several questions remain open, which encouraged research work on this topic. An example of such approach is elastic bandage compression for which pressure generation is much more complex than for socks or stockings and remains to be fully characterized. Engineering approaches were developed in this aim, providing new knowledge and understanding, and raising new questions as well, which will be presented in this section.
Deeper in the basic understanding of compression biomechanics, engineering approaches also addressed the more fundamental questions of the mechanisms actually involved in elastic compression and their effects on soft tissues of the leg. This section will present several contributions, which are mostly on the side of modeling and simulation (though directly linked to different sets of measurements) due to the complexity of performing relevant experiments on this topic. These contributions addressed various questions of the problem focusing on aspects such as the direct effect on soft tissues, on vein closure, or the development of patient-specific models to study inter-individual variability in the observed mechanisms.

**4.1. Characterizing and understanding surface pressure generation**
The clinical studies presented in section 3 evidenced the complexity of interface pressure generation, even more markedly for bandages (Clark 2003). They revealed that the law of Laplace is not sufficient in spite of being widely used in practice (Basford 2002; Schuren and Mohr 2008). In this regard, several engineering studies were proposed to better understand and characterize how interface pressure is distributed over the leg.

Experimental approaches were developed using other techniques than those presented in section 3 for clinical use. Using a measurement device previously developed in (V. Blazek and Schultz-Ehrenburg 1997), the heterogeneity of the interface pressure along the circumference of a leg cross-section was quantified showing marked variations in this pressure profile (Gaied, Drapier, and Lun 2006) (see Figure 6). Though it was performed on a rigid wooden leg, this type of measurement revealed the influence of the patient's morphology and challenged again the relevance of the law of Laplace from the perspective of individualizing compressive treatments.
Today an improvement of the experimental characterization of interface pressure faces several issues which limit novel results:
- Sensors should measure a surface distribution, not just a local or averaged measurement. Current matrix sensors have not demonstrated their reliability on complex and highly-curved surfaces like the leg. (Baldoli et al. 2016) (Rébecca Bonnaire et al. 2014)
- The influence of the sensor size deserves attention. In addition to possible averaging effects, it was shown that its geometry can alter the true pressure generated by a compressive textile (F. Chassagne et al. 2016).
- The deformation of a leg under the action of a compressive textile is generally not negligible. This highlights the need to perform such studies on soft phantom legs, or on real subjects/patients. Furthermore, this phenomenon may induce sensor misplacement, slippage and/or adhesion which should not impair the quality of measurement.

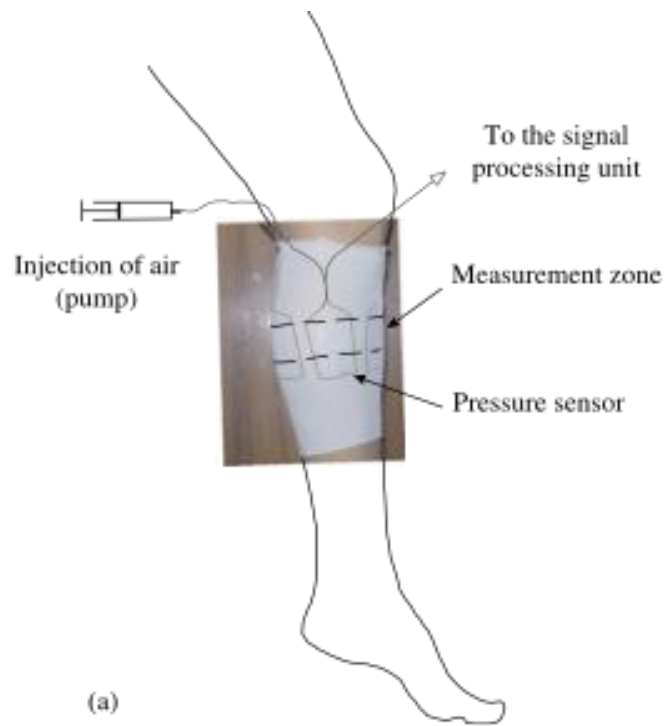

Figure 6: Surface pressure measurement device, used on a wooden leg in (Gaied, Drapier, and Lun 2006): Surface pressure measurement device, used on a wooden leg in (Gaied, Drapier, and Lun 2006)

In this context, using numerical simulation obviously turns out to be an interesting alternative. As a first step towards improved interface pressure calculation, an interesting numerical approach was presented in (Dubuis et al. 2011) where the law of Laplace was modified in order to compute interface pressure as a function of the local radius of curvature of the leg obtained from Magnetic Resonance Imaging (MRI) acquisition of the whole leg. This method – though not validated in terms of pressure values, but rather in terms of leg deformation under elastic compression – was able to assess pressure heterogeneities over the leg, hence under- or over-pressure areas which could be important from the perspectives of treatment efficacy and observance (possible link with comfort). Yet, this approach requires an MRI measurement of the leg shape which is still not easily applicable in practice. Using similar inputs, the method proposed by (X. Q. Dai et al. 2007) successfully determined interface pressure by directly modeling the motion of sock application onto the limb (see Figure 7).

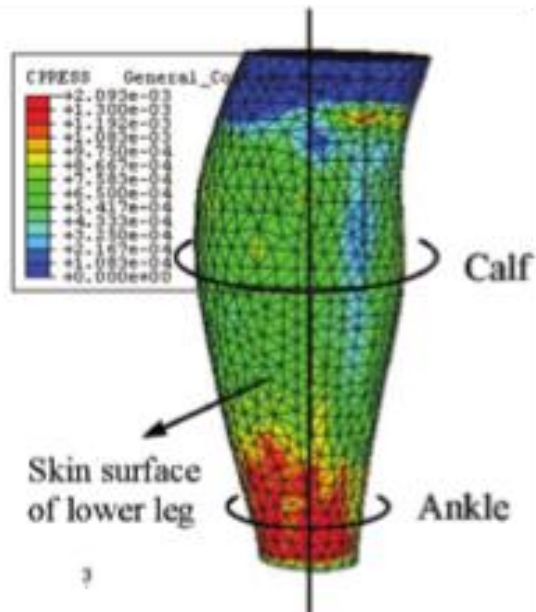

Figure 7: Skin pressure distribution obtained from a finite element simulation of a stocking being put on (X. Q. Dai et al. 2007)

The above studies were performed with compression socks, which consist in a "simple" device as it is made of a single textile layer whose stretch can be properly assessed. When studying compression bandages, complexity significantly increases as the textile layer undergoes superimposition with itself or other bandages. At least, additional frictional effects between the layers or with skin are involved in pressure generation with such bandages.

Clinical studies were numerous (see section 3) on studying this complex configuration where each bandage can also have its own properties. Likewise, part of the research effort was dedicated to addressing this more complex case. Among the most recent works, numerical models derived from the modeling technics used for socks/stockings, first showed the need for refined approaches in which friction and application process (i.e. bandage wrapping) must be considered (F. Chassagne et al. 2016). Therefore, they led to the development of a specific modeling and simulation approach in which the actual wrapping movement was simulated in order to properly integrate the effects of friction on the skin and between layers (Fanette Chassagne et al. 2018) (see Figure 8). It is interesting to note that the current technological limitations (simulation time) led the authors to develop a model reduction approach, enabling a fast assessment of pressure on any 3D leg geometry.

The main finding regarding interface pressure generation was the respective influence of parameters possibly involved in the mechanism of pressure generation. First, though their deformability must be taken into account, the elastic properties of soft tissues have little influence when their values are in the range of those measured *in vivo*. It is important to notice that this claim is only valid for interface pressure. Indeed, these properties influence the transmission of mechanical efforts through soft tissues (P.-Y. Rohan et al. 2015). Second, bandage-to-bandage friction was also a low-influence parameter, because the frictional properties (obtained from experimental measurements) were in a range that prevented any sliding between layers in this purely static case (Coulomb fiction coefficient from 0.5 to 0.7). This statement may be different when considering dynamic conditions, but this remains to be explored. Finally skin-to-bandage friction was found to be very sensitive, because bandages would tend to slide down, and then to unstretch if these frictional effects are too low. Obviously, bandage tension was the most influencing parameter overall.

In spite of these advances, several aspects that may influence interface pressure have not been fully characterized yet. Among them, short-term (period of a walking cycle) and long-term (period of a day) dynamic effects were not detailed. It is likely that slippage effects occur over time but this has not been characterized yet; it is worth reminding that the objective is a decrease of the limb volume, which should have an incidence on the medical device tension and pressure (Damstra, Brouwer, and Partsch 2008). In addition to frictional effects and possible stabilization over time, it is sensible to expect effects of hygrometry/hydration variations which could influence the mechanics of the compressive textile as well as its mechanical/biological interactions with skin (Gerhardt et al. 2008). The latter direction of research would indirectly address both the mechanical efficacy of the treatment, and comfort which is supposed to be a key to improve observance.

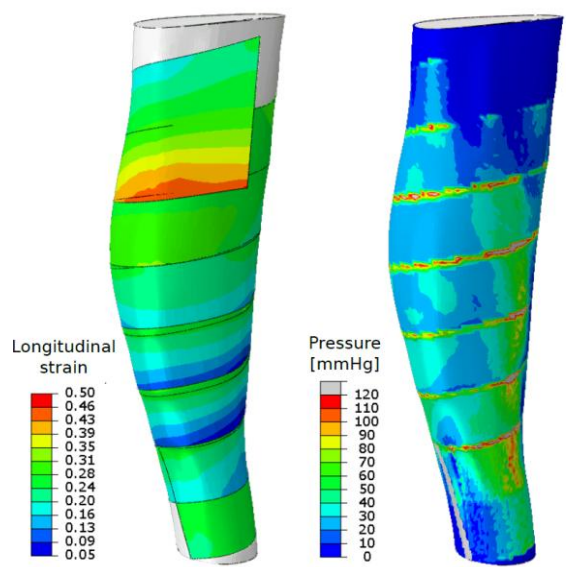

Figure 8: Finite element simulation of bandage wrapping (left: bandage longitudinal strain, right: interface pressure)

## 4.2. Effects on soft tissues of the leg

Because one of the main supposed effects of compression is a mechanical effect on the underlying soft tissues, an important effort of research was focused on this aspect, addressing the effects on soft tissues (adipose and muscular tissues) and the effect on veins which are *in fine* directly related to venous return intensity, the main objective of compression treatments.

Though venous hemodynamics improvement is the target, it is almost surprising to notice that literature is very scarce on this topic. Based on acquisitions of MR imaging and pulsed echo-Doppler data with and without elastic compression stockings, the hemodynamics effects of elastic compression were assessed via the computation of wall shear stresses in one superficial (great saphenous) vein and one deep (peroneal or posterior tibial) vein (Downie et al. 2008) (see Figure 9). This study showed that stockings were effective in raising venous wall shear stresses in prone subjects, with a much more pronounced effect in deep vessels. The latter claim may be explained by the fact that deep veins collect the resulting effects from all superficial/perforating veins and microvessels. Similar data were used by the same group to compute the reduction of vein caliber using a finite element model fed from MRI data (Wang et al. 2013). The authors also mentioned that tissue compressibility was a key parameter in this phenomenon of vein area reduction, suggesting that the effect of compression on different subjects could be very different if mechanical properties of tissues are different.

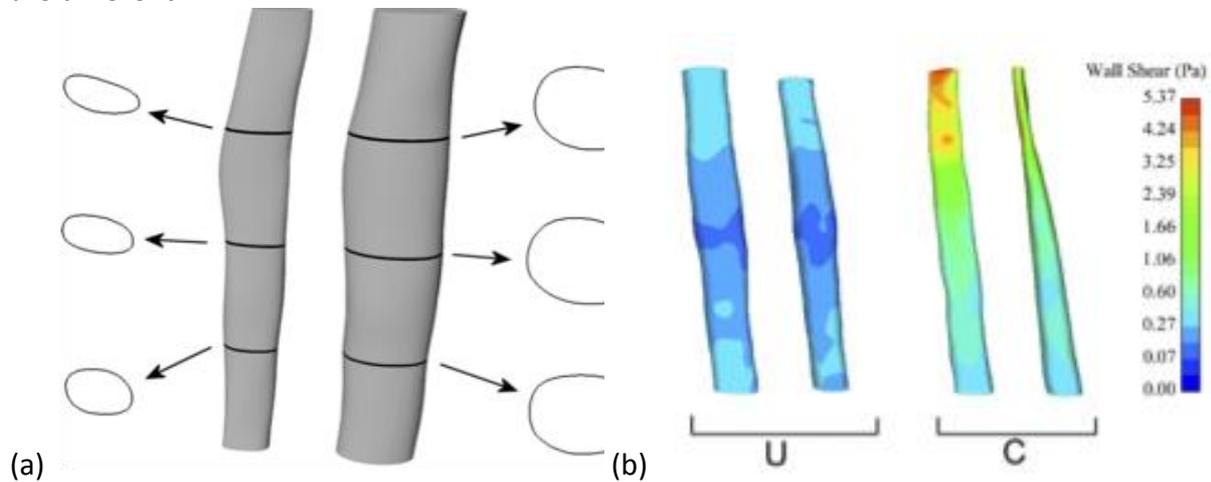

(a)  (b)

Figure 9: MRI-based reconstruction of compressed and uncompressed veins (a) and wall shear stress calculation (b) (Downie et al. 2008)

Implementing relevant experimental approaches for this question of the mechanical effects of compression on internal soft tissues is a challenge. Apart from the previously mentioned studies which mainly address veins and not other soft tissues, other works are very scarce in the literature, to our knowledge. This statement probably arises from the complexity to setup experimental studies which would require measuring internal mechanical quantities in opaque, highly-deformable materials. If *in vivo* measurements could be implemented with MR elastography, manufacturing phantoms is another alternative that faces the issue of being representative enough of the real problem.

As a consequence, the biomechanical effects of elastic compression on underlying soft tissues were addressed mainly with numerical approaches, most of which are based on finite element simulations.
Targeting this objective, two-dimensional (cross-sectional) models were developed as a first attempt to study the effects in superficial adipose tissue and deep muscular tissues. Among them, (Avril et al. 2010) developed a model using MRI data from one patient for the geometry, in which each type of tissue (adipose and muscular) was considered as homogeneous. The application of elastic compression was modeled using the law of Laplace taking the local curvature into account, and enabled computing the whole pressure field in the soft tissues. Discrepancies of more than 35% could be observed from one location to another, showing that the same compression garment may not be suitable for treating deficiencies of both the deep venous system and the large superficial veins (see Figure 10 - a). Moreover, it was shown that the internal morphology of the human leg can play an important role, in particular when considering the ratio of adipose to muscular tissue.

Towards more complexity, (G. Dai, Gertler, and Kamm 1999; Narracott et al. 2009; Wang et al. 2013) introduced veins in their 2D models in order to directly evaluate the effects on the venous network, target of the treatment. The caliber reduction of one or several veins could be quantified. Though useful from a didactic point of view, several assumptions were made such as a uniform interface pressure, homogeneous mechanical properties of tissues, which were found to be source of major sensitivity in the above-mentioned papers. One more recent work addressed these aspects and proposed two patient-specific models obtained from MR and ultrasound images to study the effect of medical compression stockings on vein closure, both for superficial and deep veins (P.-Y. Rohan et al. 2015; C. .-Y. Rohan et al. 2013). The effect on superficial varicose veins was quantified using the venous transmural pressure variation induced by the compression garment, which was found to decrease as a result of the increase of adipose tissue pressure. The transmural pressure is known to be related to the tension of the vein wall according to the law of Laplace and probably to alterations in the vein wall. Another consequence is a greater stability with respect to axial buckling and tortuosity development (Martinez et al. 2010; Badel, Rohan, and Avril 2013). In the second model where muscle compartments, the crural fascia, and deep veins were included, the combined effects of the elastic compression device and muscular contraction were investigated (see Figure 10 - b). The contribution of the compression stocking to the deep vein diameter reduction was rather small, and in fact negligible, compared to that of the contracting muscle: 3 and 9% decrease in the vein cross-sectional area were obtained with a grade II compression stocking in the supine and standing positions respectively, while complete collapse was obtained at the end of muscle activation.

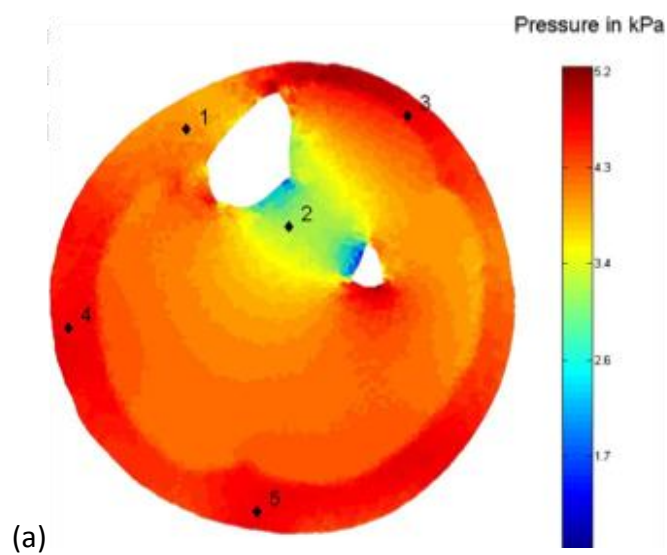

(a)

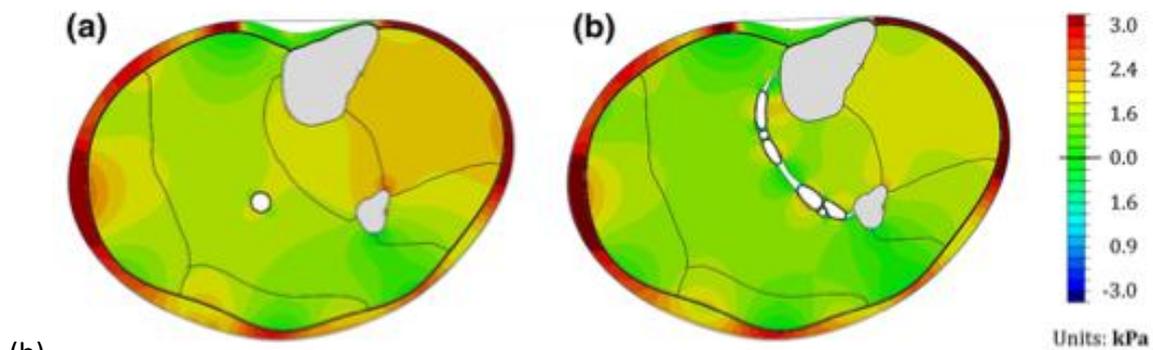

Figure 10: Hydrostatic pressure distribution in two leg cross sections. (a) Homogeneous model (Avril et al. 2010) and (b) models including compartments, fasciae and some veins (P.-Y. Rohan et al. 2015)

An approach should be mentioned which aimed at developing a 3D patient-specific model of a complete leg under elastic compression (Dubuis et al. 2011). This model was (i) patient-specific in geometry (MRI acquisition), (ii) contained two types of soft tissues, sub-cutaneous adipose tissue and muscular tissue, whose properties were also patient-specific (the approach is detailed in the following section), and (iii) the applied pressure was computed using a local law of Laplace. The results of this model confirmed the important heterogeneity of the pressure field in the internal soft tissues, hence different veins being subjected to different pressure levels. They also showed, via a study on 6 subjects, a very large inter-subject variability with, for example, variation about 50% of the pressure computed in the area of deep veins in spite of an identical grade compressive sock.

The 3D analysis of the internal pressure field is crucial to detail the actual pressure gradients induced by the elastic compression, hence the possible fluid flows, their direction and their intensity. The development of such models represents an interesting direction, in particular if the main (superficial, perforating, deep) veins and the main different components (muscle, fasciae) could be included. For now, this perspective is a challenge from a technical point of view, as a finite element mesh of a structure like this would contain a huge number of elements to properly discretize the geometry of the venous network. The question of the accuracy of imaging data required for that purpose, and reasonable acquisition conditions of these, may also be raised. Poro-elastic modeling is another alternative which has to be mentioned here.

All works above were performed in the frame of studies on various external compression devices, the question of interface pressure generation being not the main problem (as shown in section 3, this question is a topic of research in itself). Yet, the effects of morphology and mechanical properties of internal tissues were highlighted, and quantitative studies on vein closure were proposed. These works brought significant advances in the current knowledge of the mechanisms involved in compression therapy, and more specifically how the interface pressure applied by the compressive treatment impacts the mechanical state of internal tissues and structure.

### 4.3. Patient-specific approaches

In previous paragraphs, several approaches were developed on a patient-specific basis. This aspect is crucial in understanding the variability that can be observed in terms of targeted effects. The present section aims at presenting how patient-specificity was addressed in this field.

The first, and probably easiest, element of a patient-specific model that can be addressed is geometry. Indeed, current technics which can be used for that purpose are numerous and various. Furthermore, many of them are already widely used in the medical sector. They allow a wide range of scales, accuracy of details, ease of use, ease of access. Eventually the choice of a geometry acquisition technic often consists of a trade-off between ease of implementation and anatomical accuracy/detail requirements.
The richest imaging modalities are probably MRI which provides a detailed description of all tissues, and in a lesser extent Computed Tomography (CT) scanner. Access to these modalities may be difficult as it is limited by other clinical demands, the irradiating nature of CT scanner and the need to obtain authorization from competent institutions. In addition, it is important to notice that such equipment only provide supine position geometries (standing position equipment being very rare). Much simpler to use and with easier access, ultrasound echography enables to image local details but not a complete geometry.
The principle to create a geometric model from the obtained data is classical. It consists in segmenting the images and meshing the obtained result. In practice, image quality and resolution are a challenge for automatic methods, which most often led to developing semi-automatic methods (Narracott et al. 2009; Wang et al. 2013; Dubuis et al. 2012).
Multi-modality methods are an interesting alternative which was used in (C. .-Y. Rohan et al. 2013) to merge the advantage of MRI and Ultra-Sound (US) echography. The former was suitable to segment the exterior surface and internal tissues (bones, adipose tissue, muscular compartments, vessel locations) and the latter was suitable to obtain details like vein caliber and their precise geometry. Such approach requires image registration methods which are typically based on landmark registration.
Another approach presented in (Frauziols et al. 2015) was proposed to obtain a complete 2D cross-section of a leg taking advantage of the ease of use/access of US echography. This approach is based on a sequential acquisition of several US-echo images along the circumference of the leg using a specific mounting device (see Figure 10), and a registration method including the minimization of landmark-based criteria. Nevertheless, this approach is not able to provide deep geometric details.
It must be acknowledged that all previous methods are not suitable for use in a daily clinical routine. Accordingly, when only interface pressure is of interest, the external surface may be sufficient and the whole leg surface should be acquired. In this context, the use of a 3D optical scanner turns out to be a relevant alternative: it is non-irradiating, commercially available, and very fast to obtain the surface of a leg in any position. Such a device was used in a previous (clinical) study, thanks to its versatility (F. Chassagne 2017). It should be noted that assumptions on the internal structures/morphology must be made, though their sensitivity is low with respect to interface pressure.

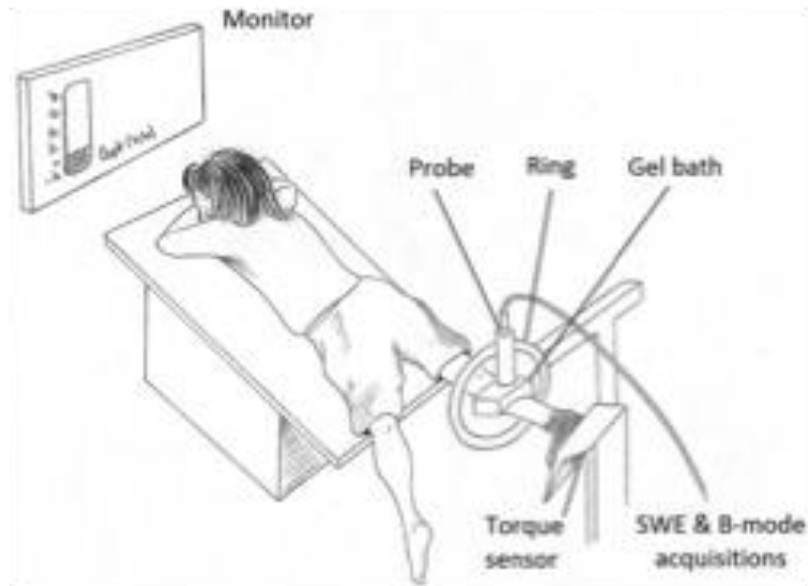

Figure 11: Experimental setup developed in (Frauziols et al. 2015): the probe is mounted on a ring, which allows collecting images around the leg. The foot is fixed through a shoe to a torque sensor. A feedback of the torque is given to the subject with the monitor. Data, including B-mode scans and SWE maps are acquired with Aixplorer (SuperSonic Imagine, Aix-en-Provence, France)

The methods to ensure geometrically patient-specific models are numerous and their implementation relies on the compromise between available equipment, and targeted objectives. With regards to patient-specific material properties for the internal soft tissues, approaches and methods are not as straight-forward. Again, the complexity of the considered method should be adjusted to the targeted objective. Most of the models presented above used linear or quasi-linear (hyper-)elastic constitutive equations. The problematics then turned into (i) defining the number of different constituents, and (ii) identifying the parameters of the material models for each of them.

For instance, in studies focusing on interface pressure characterization, all soft tissues were considered as a single homogeneous material, whereas complexity could be increased by distinguishing adipose and muscular tissue when focusing on internal mechanical fields, or even including other heterogeneities such as fasciae, distinct muscle compartments, veins and arteries when focusing on the specific response of constituent like veins.

Various approaches were developed to identify the constitutive parameters of the constituents involved in those models. Identifying a single homogeneous material for the leg soft tissues is possible thanks to external testing devices. Indentation, or localized compression, of the calf was used in several studies to this (Vannah and Childress 1996; Frauziols et al. 2016; Sengeh et al. 2016). In these studies, the parameters were identified using an inverse method based on a finite element model of the test on the patient-specific leg. The relative simplicity of this approach makes it a good candidate for clinical applications (F. Chassagne 2017). Other mechanical tests were used to identify these mechanical properties. In particular, imaging the deformation induced by the compressive stocking or cuff is a way to obtain these data. In this manner, MR images with/without compression were used as input for inverse (or similar) approaches in (Avril et al. 2010; Dubuis et al. 2013; Narracott et al. 2009). A similar approach (with/without compression) was implemented using US-echo imaging of superficial varicose veins to identify their tangent elastic moduli in standing and supine positions (C. .-Y. Rohan et al. 2013). Venous tissue behaving very non-linearly, the latter distinction was considered essential in this work.
Recently, shear wave elastography has become more and more used to assess soft tissue elastic moduli. Though implementing this technique to obtain a complete map of stiffness in the leg is complex, a method was proposed in this aim in (Frauziols et al. 2015). In addition to the underlying assumptions (for moduli reconstruction in anisotropic muscular tissue, with several heterogeneities), the applicability of this technic in a clinical context remains an issue.

Implementing patient-specific approaches mainly relies upon gathering geometric and mechanical properties. Simple as well as complex methods were proposed in the existing literature. Yet, not all of them are applicable in clinical routine, and important data are still not properly explored like fasciae mechanical properties and initial tension which may play a significant role. Other characteristics were not addressed in this section, like arterial pressure (easily measured and extrapolated at the location of interest, if the patient is at rest), muscle activity, skin friction properties, and reflex contractions which could be influenced via proprioceptive effects, for instance. These would require other equipment that were not used, to our knowledge, for consideration in biomechanical studies of elastic compression.
All presented studies and approaches were useful in understanding trends in the effects of elastic compression that are directly related to the inter-individual variability measured in these studies. Among interesting conclusions that could be drawn, it is worth noting two examples of them. First, the mechanical properties (stiffness) of calf soft tissue have negligible influence on the interface pressure value, provided they are in a reasonable range, i.e. deformable. Second, geometry and tissue mechanical properties have strong (spatial) impact on the internal distribution of pressure, and in particular on the effect on veins.
Such conclusions are interesting for clinical practice and will potentially have implications in designing specific care strategies. The objective of the following section is to specifically address and detail new knowledge and results that were obtained from clinical and research studies, along with their possible implications for the clinical practice.

**5 – Contribution to the clinical practice**
The present section sums up the essential findings and new relevant knowledge arising from the previously presented engineering and clinical research works.

Regarding internal mechanisms of elastic compression, which can be related to treatment efficacy, the following can be retained:
- Though they have little effect on interface pressure, the mechanical properties of soft tissues (e.g. elasticity) influence pressure transmission to veins and internal soft tissues.
- The internal distribution of pressure in soft tissues is also influenced by the external and internal geometry of the leg. In particular, the pressure observed at the location of deep veins may be very different from one patient to another.
- When focusing on the effects of compression on veins, it can be summarized that the response of superficial veins to compression is subject to three main factors: the vein size, the local radius of curvature of the leg and the fat stiffness. On the contrary, although the effect on deep veins may be noticeable in supine position, previous studies showed that, in standing position, the effect of compression is marginal compared to that of muscular contraction.

Interface pressure is on the one hand the condition to obtain the expected internal effects, but on the other hand it may compromise the compliance to the treatment. Though the latter is not addressed, the following should be mentioned about interface pressure:
- Bandage application is a sensitive process which highly depends on the operator and the application technique. As a consequence, a marked variability may be observed in the interface pressure applied onto skin and this should be carefully considered. Improving the control of bandage stretch and application technique are probably the keys to a better control of interface pressure.
- Bandage to skin friction plays a key role in compression by bandages. Indeed, it is essential in maintaining the wrapped bandage and preventing it from falling down. Its effect on pressure generation remains to be clarified, as the law of Laplace is questioned in this context.
- For compression stockings or socks, the stiffness of the sock, the size of the sock and the size of the leg are the main parameters influencing interface pressure. They are also the parameters used for choosing the appropriate treatment. However, studies presented in section 4 highlighted different pressure distribution patterns for different morphologies. In particular, very homogeneous pressure distributions are observed in cases of round leg cross-sections whereas high pressure spots may be found in other cases. This additional aspect is usually not considered though it may have important impact on comfort, hence observance of the treatment. It is important to note that these results were obtained for purely static simulation and should be confirmed in dynamic conditions.

Beyond the understanding of elastic compression mechanisms, it is believed that these findings can help enhance care management.
An important impact of these research-level results would be to head towards an improved personalization of compression treatment. The complete external geometry of the leg could be used to estimate personalized interface pressure distributions and improve the choice of the right size of a device. In the longer term, it is believed that it will be possible to assess the impact of a treatment on the venous return which would help in targeting the right treatment. For instance, the influence of bandage application order – in the case of multi-layer bandages – was highlighted and this specificity would deserve to be considered.
The presented results also underline the needs to standardize practices among healthcare professional. For example, specific or improved training of practitioners willing to apply compression bandages could be encouraged. On the same aspect, bandage design could be addressed for a better control of application.

## 5 - Prospects
As already mentioned, clinical studies gave limited results, but could at least conclude that the maximum applied pressure should not exceed 70 mmHg in standing position and that training medical staff is necessary to reduce complications related to over-pressure; the correlation between phantom and real legs is established in the case of static pressure measurement.

Clinical studies opened the way to more questions that are important perspectives in this research area.
Firstly, the applied pressure varies with time. Several authors studied the time-dependence of the applied pressure to the knitting process, the material viscoelasticity, the leg volume variation or the patient activity, but these studies considered only one parameter at a time and it is still impossible to rank these phenomena or even consider the effects of interactions among them. No numerical approach could describe time-dependent phenomena yet. Of course, compression level is crucial for the efficacy of the treatment, therefore the evolution of the tension or the leg volume with time and/or patient activity is a major prospect.
Indeed, from a clinical point of view, such prospect cannot be achieved in a short term: a multi-parametric study should lead to a too large cohort, and unwanted variabilities related both to patients and operators should make variations statistically non-significant. A lot of work is still to be done to extract the relevant parameters from the mechanical behavior of the textiles and from the patient daily life activity, and also to reduce the inter-operator variability.
From a numerical modeling point of view also, studying the time-dependence of hosiery compression is very ambitious; it should include new developments on the textile behavior modeling and on the fluid-structure interactions in the leg, maybe using poro-elastic modeling. Blood return is also dependent on active processes. The muscular contraction in the calf and in the foot should be put in a dynamic biomechanical model of the limb, including the foot.
For now, computer time has been a limitation and innovative numerical methods should be found; because modeling the hosiery to leg contact is computationally expensive, one way to reduce the global computing time should be the development of an analytical contact model based on the law of Laplace and including the tangential effects (friction or adhesion).

Second, as it has already been mentioned, the numerical models could describe the mechanisms of action of the hosiery onto the veins. Even if the applied pressure poorly depends on the soft tissue properties, these properties have a great influence on the load transfer from the skin to the veins, so another limitation lies in the knowledge of the patient-specific soft tissue properties (skin, fat, muscles, aponeurosis), and the mechanical interaction between them. Therefore, the second prospect proposed here should be the development of experimental methods to quantify relevant material properties of the leg. These methods should ideally be available in the clinical context in order to evaluate the efficiency of a treatment for a specific patient.

Last, the skin-textile interface has been poorly studied in the context of compression therapies. Skin, which is the natural mechanical, chemical and thermal protection of the body is highly stressed by compression or contention devices and it might be the cause of the low level of compliance to the compression-based treatments.
Two main risks are related to the skin: a high compression, even during short times can lead to ulcers, and irritation should be related to skin shear. The latter has been partially addressed by medical doctors but studies from biophysicists all focused on ulcers or prosthesis complications. The study of the skin-textile interface in the context of compression therapies should be developed both from medical and engineering angles of view. Clinical studies should be designed to establish the conditions leading to irritation. On the other hand, experimental studies should better describe the skin behavior under pressure and shear according to various parameters of the skin (moisture, hydration …), the hosiery (textile microstructure, roughness…) or the interaction of both (load, shear distance …) in order to understand basic mechanisms leading to skin diseases and to prevent them.